\documentclass[%
 reprint,
superscriptaddress,
 amsmath,amssymb,
 aps,
pra,
floatfix,
]{revtex4-1}

\usepackage{graphicx}
\usepackage{dcolumn}
\usepackage{bm}

\begin{document}

\preprint{APS/123-QED}

\title{The clamped intensity of femtosecond laser pulses varying with gas pressure in the presence of external focusing}

\author{Quanjun Wang}
\author{Yuxuan Zhang}
\author{Yue Zheng}
\author{Zhoumingyang Zhu}
\author{Pengji Ding}
\thanks{dingpj@lzu.edu.cn}
\author{Zuoye Liu}
\thanks{zyl@lzu.edu.cn}
\author{Bitao Hu}
\affiliation{School of Nuclear Science and Technology, Lanzhou University, Lanzhou 730000, People's Republic of China}

\date{\today}

\begin{abstract}
We perform a theoretical investigation of the clamped laser intensity inside the filament plasma as a function of gas pressure with external focusing. Unlike the clamped intensity under the self-focusing condition, which is independent on the gas pressure, the clamped intensity with external focusing decreases with the gas pressure. Our findings can explain the changes of the signals of femtosecond-laser-induced 391-nm forward emission and fluorescence with the nitrogen gas pressure.

\end{abstract}

\pacs{Valid PACS appear here}
\maketitle


Femtosecond laser pulses propagating in gases with the power larger than a critical power $P_{\mathrm{cr}}$ produce self-guided high-intensity plasma filaments~\cite{marburger1975self,couairon2007femtosecond,chin2010femtosecond}. Because of a balance between Kerr self-focusing and plasma defocusing, the laser intensity is clamped inside the filaments~\cite{chin2010femtosecond,liu2014intensity}. The clamped intensity $I_{\mathrm{c}}$ sets an upper limit to the intensity at the self-focus in gases. Once the clamped laser intensity is achieved, it is not only stabilized along the propagation distance, but also almost invariable with the increase of the input laser energy. The intensity clamping is one of the fundamental characteristics of the filamentation phenomenon. It governs the major dynamics of the laser-gas interaction and is a key to understand various physical phenomenon inside the filament, such as tunnel ionization~\cite{chin2016tunnel}, plasma fluorescence~\cite{becker2001intensity}, the cut-off frequency of the high order harmonic spectrum~\cite{lange1998high}. The self-focusing clamped intensity in air is $\sim 4\times 10^{13}~\mathrm{W/cm^2}$~\cite{kasparian2000critical}.  

In many experiments, an external lens is used to force self-focusing within the limit of the length of the medium and the laboratory space. Under external focusing, the intensity clamping still holds but the clamped intensity increases~\cite{theberge2007self,kosareva2009can,xu2012intensity,liu2010tightly}. Bernhardt {\it et al} showed that the peak intensities inside filaments were experimentally determined to be about $6.4\times10^{13}~\mathrm{W/cm^2}$ and $1.7\times10^{14}~\mathrm{W/cm^2}$ by using a 20-cm focal-length lens in air and argon, respectively~\cite{xu2012intensity}. With a tighter external focusing ($\mathrm{f=12.7~cm}$), Liu {\it et al} have found that the peak intensity is clamped at a level of $6\times10^{14}~\mathrm{W/cm^2}$ by measuring the electron density in air~\cite{liu2010tightly}.

The theoretical analysis showed that the clamped intensity is independent on gas pressure under self-focusing condition~\cite{chin2010femtosecond}. Is this claim also established with external focusing? This article answers the question. 

We begin with the balance of the nonlinear index of refraction between Kerr self-focusing $\Delta n_{\mathrm{Kerr}}$ and plasma defocusing $\Delta n_{\mathrm{p}}$. $\Delta n_{\mathrm{Kerr}}$ is
\begin{equation}
    \Delta n_{\mathrm{Kerr}}=n_2I,
\end{equation}
where $n_2$ is the Kerr nonlinear index of refraction and $I$ is the laser intensity. $\Delta n_{\mathrm{p}}$ is 
\begin{equation}
    \Delta n_{\mathrm{p}}=\frac{N_{\mathrm{e}}(I)}{2N_{\mathrm{crit}}},
\end{equation}
where $N_\mathrm{e}$ and $N_\mathrm{crit}$ are the electron density and critical plasma density, respectively. $N_\mathrm{crit}=\frac{\varepsilon_0 m \omega^2}{e^2}$ with $\varepsilon_0$, $m$, $\omega$ and $e$ denoting the vacuum permittivity, the mass of electron, the laser angular frequency and the elementary charge, respectively. For a titanium–sapphire laser with a central wavelength of 800 nm, $N_\mathrm{crit}$ is $1.7\times 10^{21}~\mathrm{cm^{-3}}$. The balance equation reads
\begin{equation}
    n_2I=\frac{N_\mathrm{e}(I)}{2N_\mathrm{crit}}.
    \label{self_focus_balance}
\end{equation}
$n_2=\kappa N_\mathrm{gas}$ is linearly proportional to the neutral gas density $N_\mathrm{gas}$ with $\kappa=1.88\times 10^{-38}~\mathrm{cm^5/W}$ for nitrogen gas~\cite{borzsonyi2010measurement,nibbering1997determination}. The electron density $N_\mathrm{e}=R(I) \tau_\mathrm{I} N_\mathrm{gas}$ is also linearly proportional to the neutral gas density $N_\mathrm{gas}$. $R(I)$ and $\tau_\mathrm{I}$ represent the ionization rate and ionization time. The Eq.~\ref{self_focus_balance} can be written as 
\begin{equation}
    \kappa I=\frac{R(I)\tau_\mathrm{I}}{2N_\mathrm{crit}}.
    \label{self_focus_clamped}
\end{equation}
It can be seen the clamped intensity under self-focusing is unchanged with the gas pressure. 

\begin{figure}[ht]
\centering
\includegraphics[width=3.4 in]{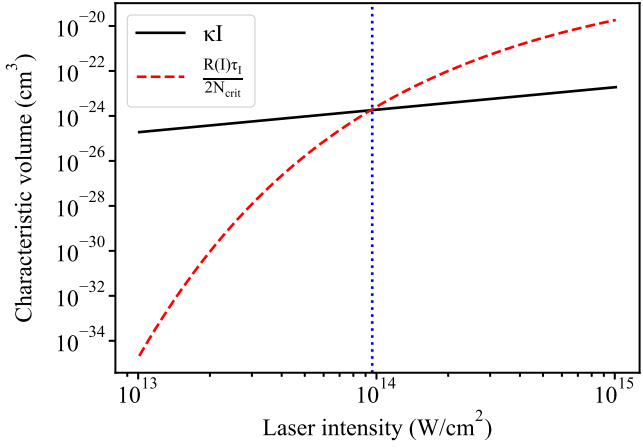}
\caption{The characteristic volume $\kappa I$ and $\frac{R(I)\tau_\mathrm{I}}{2N_\mathrm{crit}}$ as a function of 800-nm laser intensity. The ionization rate of nitrogen is calculated by ADK model.}
\label{I_cr_self-focus_ADK}
\end{figure}

Choosing the gas as nitrogen, the ionization rate $R(I)$ is calculated by Ammosov, Delone and Krainov's (ADK) model~\cite{ammosov1986tunnel}. The lowest three ionic states $\mathrm{N_2^+}(X^2\Sigma_g^+$), $\mathrm{N_2^+}(A^2\Pi_u$) and $\mathrm{N_2^+}(B^2\Sigma_u^+$) with corresponding ionization potentials 15.581, 16.699 and 18.875 eV~\cite{itikawa2006cross}, respectively, are taken into account, as all of them have been observed in femtosecond laser fields~\cite{gibson1991dynamics}. The ionization time $\tau_\mathrm{I}$ is set to be 100 fs which is on the order of the laser pulse duration. We can draw the right-hand side $\frac{R(I)\tau_\mathrm{I}}{2N_\mathrm{crit}}$ of Eq.~\ref{self_focus_clamped} as a function of laser intensity as shown in Fig.~\ref{I_cr_self-focus_ADK} (dashed red line). The left-hand side $\kappa I$ of Eq.~\ref{self_focus_clamped} is easily obtained as shown in Fig.~\ref{I_cr_self-focus_ADK} (solid black line). As $\kappa I$ and $\frac{R(I)\tau_\mathrm{I}}{2N_\mathrm{crit}}$ have the unit of cubic meter, they are called characteristic volumes here. The abscissa of the intersection of the two curves is the self-focusing clamped intensity. The $I_{\mathrm{c}}$ is $9.6\times 10^{13}~\mathrm{W/cm^2}$ in nitrogen for the 800-nm femtosecond laser pulse. 

\begin{figure}[ht]
\centering
\includegraphics[width=3.4 in]{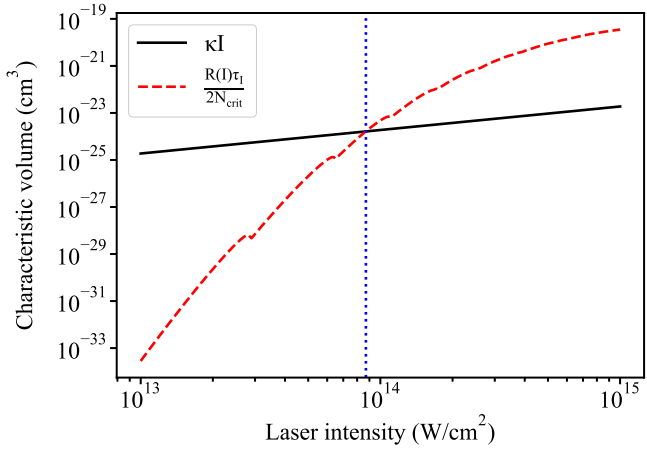}
\caption{The characteristic volume $\kappa I$ and $\frac{R(I)\tau_\mathrm{I}}{2N_\mathrm{crit}}$ as a function of 800-nm laser intensity. The ionization rate of nitrogen is calculated by PPT model.}
\label{I_cr_self-focus_PPT}
\end{figure}

In the range of $10^{13}-10^{14}~\mathrm{W/cm^2}$ of 800-nm laser, another model of Perelemov, Popov and Trentev’s (PPT) model is often used to calculate the ionization rate, which fits very well the experimental ion yields~\cite{perelomov1966ionization,talebpour1998suppressed,talebpour1999semi}. Figure~\ref{I_cr_self-focus_PPT} shows that the self-focusing clamped intensity is $8.7\times 10^{13}~\mathrm{W/cm^2}$ in nitrogen with the ionization rate calculated by PPT model. The simulated clamped intensities with the helps of ADK model and PPT model are closed to each other. They are higher than the value of $4\times 10^{13}~\mathrm{W/cm^2}$ estimated by Kasparian {\it et al} in air, because the ionization potential of nitrogen is higher than that of oxygen which accounts for more than 80\% of the overall plasma in air~\cite{kasparian2000critical}. 

\begin{figure}[ht]
\centering
\includegraphics[width=3.4 in]{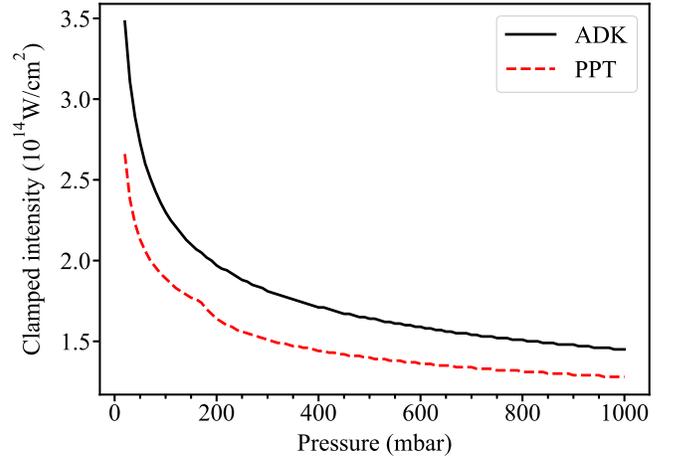}
\caption{The clamped intensity as a function of nitrogen gas pressure $p$ with the $\Delta n_{\mathrm{lens}}=10\times 4.35\times10^{-5}$. The ionization rates are calculated by ADK model and PPT model for the solid black line and the dashed red line, respectively. The relationship between $N_{\mathrm{gas}}$ and $p$ is $N_{\mathrm{gas}}=2.41\times p~(\mathrm{mbar})\times10^{16}~\mathrm{cm^{-3}}$ in the simulation.}
\label{I_cr_with_pressure}
\end{figure}

When an external lens is added, the new equilibrium becomes~\cite{chin2010femtosecond}
\begin{equation}
   \Delta n_{\mathrm{lens}} + \kappa N_{\mathrm{gas}} I=\frac{R(I)\tau_\mathrm{I} N_{\mathrm{gas}}}{2N_\mathrm{crit}},
   \label{balance_with_lens}
\end{equation}
where $\Delta n_{\mathrm{lens}}$ is the change in the index of refraction generated by the external focusing. Since the ionization rate $R(I)$ increases non-linearly with the laser intensity, there exists a larger intensity that makes the left-hand side equals right-hand side in Eq.~\ref{balance_with_lens}. Hence, the clamped intensity increases with the external focusing. We do not know how to calculate the value of $\Delta n_{\mathrm{lens}}$. Previous studies have revealed that it is related to the focal length~\cite{liu2014intensity,theberge2007self,kosareva2009can,xu2012intensity,liu2010tightly}. In the current case, it is taken as a parameter and $\Delta n_{\mathrm{lens}}=b\times 4.35\times10^{-5}$ is used, where $b$ is a coefficient and $4.35\times10^{-5}$ is the value of $\Delta n_{\mathrm{Kerr}}$ with the self-focusing clamped intensity $9.6\times 10^{13}~\mathrm{W/cm^2}$ in 1000-mbar nitrogen gas. With the help of Eq.~\ref{balance_with_lens}, we can explore the clamped intensity as a function of nitrogen pressure $p$ ($N_{\mathrm{gas}}$). Figure~\ref{I_cr_with_pressure} shows that the simulated $I_{\mathrm{c}}$ decreases with the nitrogen gas pressure from 10 to 1000~mbar. The $I_{\mathrm{c}}$ drops significantly fast at the low pressure range (10-100~mbar) and decreases slowly at high pressures. This feature is due to that the clamped intensity varies inversely with the pressure according to Eq.~\ref{balance_with_lens}. The clamped intensities are more than $1\times 10^{14}~\mathrm{W/cm^2}$ with the coefficient $b=10$ by using both ADK and PPT models to calculate the ionization rate. The $I_{\mathrm{c}}$ obtained via ADK model is higher than that obtained via PPT model, because the ionization predicted by ADK model is smaller than that predicted by PPT model for the same laser intensity~\cite{talebpour1999semi}.

Two examples related to the 391-nm transition from $\mathrm{N_2^+}(B^2\Sigma_u^+,\nu'=0)$ to $\mathrm{N_2^+}(X^2\Sigma_g^+,\nu=0)$ are given here, which can be explained by the simulated results of Fig.~\ref{I_cr_with_pressure}.

\begin{figure}[ht]
\centering
\includegraphics[width=3.4 in]{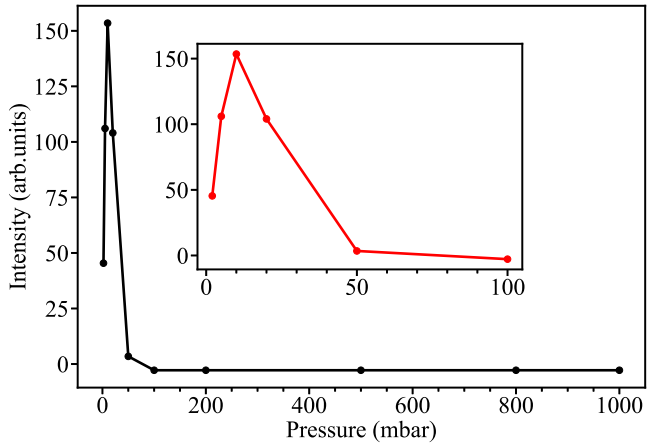}
\caption{The intensity of the 391-nm forward emission of $\mathrm{N_2^+}$ as a function of gas pressure. The 800-nm femtosecond laser (35 fs, 3 mJ and 1 kHz) is focused by a 40-cm focal-length lens in pure nitrogen. A seed pulse around 391 nm is generated by second harmonic generation in a 200-$\mu$m-thickness $\beta$-barium borate ($\beta$-BBO) crystal on the second arm. The weak seed pulse is combined with the 800-nm pump pulse by using a dichromatic mirror and also focused bt the 40-cm focal-length lens. The 391-nm forward spectral signal is collected by a high-resolution spectrometer (Model 2061, McPherson, Inc.), and the residual pump beam is removed by a 400-nm filter. The intensity is the integral of the spectral signal.}
\label{391-nm_SR_intensity_with_pressure}
\end{figure}

1. The dependence of 391-nm forward emission inside the plasma filaments on nitrogen gas pressure. The 391-nm forward emission was found by Yao {\it et al} in 2011 and has attracted a lot of attention~\cite{yao2011high,yao2013remote,yao2016population,liu2017unexpected,tikhonchuk2021theory}.
Figure~\ref{391-nm_SR_intensity_with_pressure} shows the intensity of the 391-nm forward emission as a function of gas pressure. It first goes up with the pressure increasing from 2 to 10 mbar and then declines in the range of 10--100 mbar. After the pressure exceeds 100 mbar, the 391-nm forward emission disappears. The similar results have been reported previously~\cite{wang2015population,mysyrowicz2019lasing}. Wang {\it et al} and Mysyrowicz {\it et al} observed the optimum pressures at 10 mbar and 30 mbar, respectively, by using the convex lenses of the focal length 30 cm and 40 cm~\cite{wang2015population,mysyrowicz2019lasing}. The 391-nm forward emission is confirmed to be superradiance and/or superfluorescence inside the plasma filaments pumped by 800-nm femtosecond lasers~\cite{li2014signature,liu2015recollision,wang2021superradiance}. One of the prerequisites of superradiance and/or superfluorescence is population inversion between the upper and lower energy states~\cite{dicke1954coherence,bonifacio1971quantum,macgillivray1976theory,polder1979superfluorescence,bonifacio1975cooperative}. By solving the time-dependent Schr{\"{o}}dinger equation, Xu {\it et al} calculated the population dynamics in the $\mathrm{N_2^+}$ and showed that the population inversion between $\mathrm{N_2^+}(B^2\Sigma_u^+)$ to $\mathrm{N_2^+}(X^2\Sigma_g^+)$ is established when the 800-nm laser intensity exceeds $2.2\times 10^{14}~\mathrm{W/cm^2}$~\cite{xu2015sub}. The relative population difference of $\mathrm{N_2^+}(B^2\Sigma_u^+)$ to $\mathrm{N_2^+}(X^2\Sigma_g^+)$ varies from 0 to 0.55 corresponding to the 800-nm laser intensity ranging from $2.2-4\times 10^{14}~\mathrm{W/cm^2}$~\cite{xu2015sub}.

The reduction and disappearance of the 391-nm forward emission can be explained by the decrease of clamped intensity as shown in Fig.~\ref{I_cr_with_pressure}, which leads to a weaker and weaker inversion level. When the clamped intensity is lower than a certain value, the population inversion can not be achieved and the 391-nm forward emission disappears. The increase of the 391-nm forward emission is due to the increased number of $\mathrm{N_2^+}(B^2\Sigma_u^+)$ and $\mathrm{N_2^+}(X^2\Sigma_g^+)$ with the unchanged relative population difference, by assuming that  
the laser intensities are unchanged in the range of 2-10 mbar and have the same value as the 10 mbar.

\begin{figure}[ht]
\centering
\includegraphics[width=3.4 in]{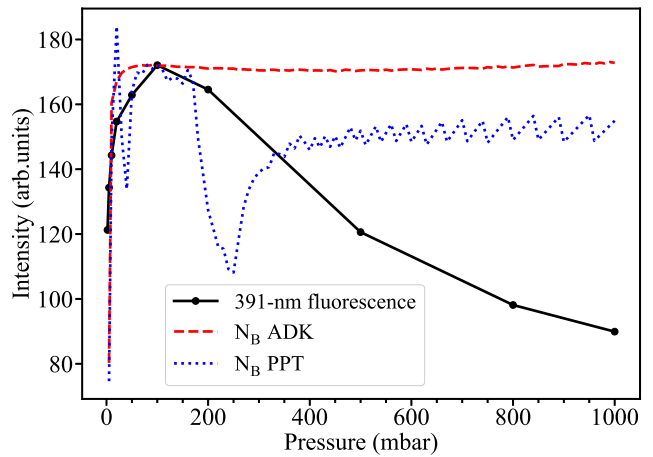}
\caption{The intensity of the 391-nm side fluorescence (solid black line) as a function of gas pressure. The 800-nm femtosecond laser (35 fs, 3 mJ and 1 kHz) is focused by a 40-cm focal-length lens in pure nitrogen. The intensity is the integral of the spectral signal. The numerical prediction of the number of $\mathrm{N_2^+}(B^2\Sigma_u^+)$ by using ADK (dashed red line) and PPT (dotted blue line) models to calculated the ionization rate.}
\label{391-nm_fluo_intensity_with_pressure}
\end{figure}

2. The dependence of 391-nm side fluorescence inside the plasma filaments on nitrogen gas pressure. 
As shown in Fig.~\ref{391-nm_fluo_intensity_with_pressure}, the fluorescence intensities first increase and then decrease with increasing the gas pressure from 2 to 1000 mbar. The similar behaviour of 391-nm fluorescence emission have been reported~\cite{wang2015population,talebpour2001spectroscopy}. Unlike the 391-nm forward emission, the fluorescence emission exists in the full range of gas pressure and the optimum pressure shows up around 100 mbar. As is well known, the fluorescence intensity is proportional to the population of the upper level and not related to the lower level. The number of the $\mathrm{N_2^+}(B^2\Sigma_u^+)$ can be expressed by 
\begin{equation}
    N_{\mathrm{B}}=R_{\mathrm{B}}(I_{\mathrm{c}})\tau_{\mathrm{I}}N_{\mathrm{gas}}V,
    \label{number_B}
\end{equation}
with $R_{\mathrm{B}}(I_{\mathrm{c}})$ and $V$ denoting the ionization rate of $\mathrm{N_2^+}(B^2\Sigma_u^+)$ and the plasma volume. According to $I_{\mathrm{c}}=\frac{E_{\mathrm{in}}}{\tau \pi r_{\mathrm{c}}^2}$ with $E_{\mathrm{in}}$ and $\tau$ being the pule energy and duration, the radius $r_{\mathrm{c}}$ of the filamemt plasma will increase with the gas pressure as the clamped intensity decreases in Fig.~\ref{I_cr_with_pressure}, By assuming that the plasma length is unchanged, the plasma volume can be written as $V=\frac{g}{I_{\mathrm{c}}}$, where $g$ is a coefficient. Substituting $V=\frac{g}{I_{\mathrm{c}}}$ into Eq.~\ref{number_B}, we get 
\begin{equation}
    N_{\mathrm{B}}=R_{\mathrm{B}}(I_{\mathrm{c}})\tau_{\mathrm{I}}N_{\mathrm{gas}}\frac{g}{I_{\mathrm{c}}}.
    \label{number_B_final}
\end{equation}

The assumption that the laser intensity in the range of 2--10 mbar has the same value as the 10 mbar is utilized again. By using the clamped intensities $I_{\mathrm{c}}$ in Fig.~\ref{I_cr_with_pressure}, the population of $\mathrm{N_2^+}(B^2\Sigma_u^+)$ is calculated based on Eq.~\ref{number_B_final}. Figure \ref{391-nm_fluo_intensity_with_pressure} shows the simulated $N_{\mathrm{B}}$ as a function of nitrogen gas pressure from 2 to 1000 mbar. The dashed red line is acquired by putting the solid black line in Fig.~\ref{I_cr_with_pressure} into Eq.~\ref{number_B_final} and using the ADK model to calculate the ionization rate of $\mathrm{N_2^+}(B^2\Sigma_u^+)$. The dotted blue line is acquired by putting the dashed red line in Fig.~\ref{I_cr_with_pressure} into Eq.~\ref{number_B_final} and using the PPT model to calculate the ionization rate of $\mathrm{N_2^+}(B^2\Sigma_u^+)$. The simulated $N_{\mathrm{B}}$ in Fig.~\ref{391-nm_fluo_intensity_with_pressure} is normalized to the fluorescence intensity at 100 mbar. The $N_{\mathrm{B}}$ (dashed red line) induced by ADK model goes up with the gas pressure increasing from 2 to 20 mbar and are almost unchanged in the range of 50--1000 mbar. The $N_{\mathrm{B}}$ (dotted blue line) induced by PPT model has similar tendency with that induced by ADK model, except the irregular oscillations in the 0--400 mbar pressure region. The irregular oscillations are due to the non-smooth ionization rate calculated by PPT model as shown in Fig.~\ref{I_cr_self-focus_PPT}. 

The intensities of 391-nm fluorescence increase at low pressures because of the increased population $N_{\mathrm{B}}$. The excited-state decay contains radiative (fluorescence) process and non-radiative processes such as collisional quenching. The constant of the rate of $\mathrm{N_2^+}(B^2\Sigma_u^+)$ quenching by $\mathrm{N_2}$ molecules is $3.0\times10^{-10}~\mathrm{cm^3/s}$~\cite{arnold2012excited,valk2010measurement,shakhatov2008kinetics}. The non-radiative decay rate caused by the collisional quenching is proportional to the gas pressure and the corresponding non-radiative decay lifetimes from 100 to 1000 mbar are 1.38 to 0.38 ns, which is larger than the fluorescent life time 62 ns~\cite{valk2010measurement}. Hence, the depopulation of 
$\mathrm{N_2^+}(B^2\Sigma_u^+)$ through the non-radiative decay is intenser and intenser with the pressure increasing and the fluorescence intensities decrease in the range of 100-1000 mbar as shown in Fig.~\ref{391-nm_fluo_intensity_with_pressure}.

To conclude, the clamped intensity inside the filament plasma as a function of gas pressure is theoretically investigated. With the external focusing, the clamped intensities decrease as the gas pressure increases. The variation of 391-nm forward emission and side fluorescence of $\mathrm{N_2^+}$ with gas pressure are explained with the help of our findings. We hope it can be used as a fundamental characteristic of filament plasma to understand more relevant phenomena.

This work was supported by the National Natural Science Foundation of China (Grants No. U1932133, No. 11905089, and No. 12004147). 

Q. Wang and Y. Zhang contributed equally to this work.

\end{document}